\documentclass[12pt]{iopart}
\usepackage[latin1]{inputenc}
\usepackage{amssymb}
\usepackage{graphicx}
\usepackage{graphics}

\begin{document}
\newcommand{\nf}{n$_f$}
\newcommand{\smd}{Sm$^{2+}$}
\newcommand{\smt}{Sm$^{3+}$}
\newcommand{\tmd}{Tm$^{2+}$}
\newcommand{\tmt}{Tm$^{3+}$}
\newcommand{\pstar}{p$^{*}$}
\newcommand{\tk}{$T_{K}$}
\newcommand{\tkl}{$T_{KL}$}
\newcommand{\tkd}{$T_{KL}^{2+}$}
\newcommand{\tkt}{$T_{KL}^{3+}$}
\newcommand{\fz}{f$^{0}$}
\newcommand{\fu}{f$^{1}$}
\newcommand{\fd}{f$^{2}$}
\newcommand{\smb}{SmB$_6$}
\newcommand{\cf}{$\Delta_{CF}$}
\newcommand{\tn}{$T_{N}$}

\title[Valence and magnetism in IVC]{Valence and magnetic ordering in intermediate valence compounds : TmSe versus SmB$_6$}

\author{J.~Derr, G.~Knebel, G.~Lapertot, B.~Salce, M-A.~M\'easson and J.~Flouquet}

\address{Département de Recherche Fondamentale sur la Matière Condensée, CEA~Grenoble, 17 rue des Martyrs, 38054 Grenoble Cedex 9, France}
\ead{julien.derr@cea.fr}
\begin{abstract}
The intermediate valent systems TmSe and \smb\ have been investigated up to 16 and 18~GPa by ac microcalorimetry with a pressure ($p$) tuning realized in situ at low temperature. 
  For TmSe, the transition from an antiferromagnetic insulator for $p<3$~GPa to an antiferromagnetic metal at higher pressure has been confirmed. A drastic change in the $p$ variation of the N\'eel temperature ($T_N$) is observed at 3~GPa. In the metallic phase ($p>3$~GPa) , \tn\ is found to increase linearly with $p$. A similar linear $p$ increase of \tn\ is observed for the quasitrivalent compound TmS which is at ambiant pressure equivalent to TmSe at $p \sim 7$~GPa.
 In the case of \smb\ long range magnetism has been detected above $p \sim 8$~GPa, i.e. at a pressure slightly higher than the pressure of the insulator to metal transition. However a homogeneous magnetic phase occurs only above 10~GPa. The magnetic and electronic properties are related to the renormalization of the 4f wavefunction either to the divalent or the trivalent configurations. As observed in SmS, long range magnetism in \smb\ occurs already far below the pressure where a trivalent \smt\ state will be reached. 
 It seems possible, to describe roughly the physical properties of the intermediate valence equilibrium by assuming formulas for the Kondo lattice temperature depending on the valence configuration. Comparison is also made with the appearance of long range magnetism in cerium and ytterbium heavy fermion compounds.
\end{abstract}

\maketitle

\section{Introduction}

\paragraph{} Recently, a major interest was the study  of the high pressure phase diagrams of heavy fermion compounds (HFC)\cite{revuedeJacques}. However, in these systems, the departure from the trivalent configuration is weak ; the occupation number \nf\ of the $4f^{1}$ is nearly one. Unusual magnetic properties found notably on ytterbium HFC such as YbRh$_2$Si$_2$ push to revisit other situations with magnetic and valence fluctuations occuring between two 4f  configurations. The cases of intermediate valence compounds (IVC) as SmS, \smb\, and TmSe, are particularely interesting\cite{Wachter}.

\begin{figure}
\begin{center}
\includegraphics[width=7cm]{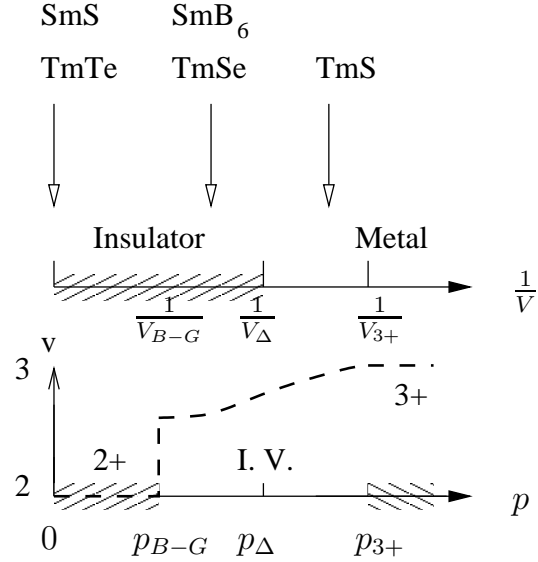}
\caption{Valence state, as a function of the density, i.e. the inverse of the molar volume $V$. For $V<V_{B-G}$, the system jumps from 2+ black (B) phase to IV gold (G) phase. up to $V_{B-G}>V>V_{\Delta}$ the system is still insulating (The dashed lines represent the insulating caracter). Magnetism looks also governed by the 2+ configuration. For $V<V_{\Delta}$, the system is metallic, and magnetism looks governed by the 3+ configuration. The trivalent limit will appear for $V_{3+}<V_{\Delta}$. Long range magnetism in SmS and \smb\ appear for $\frac{1}{V}\sim \frac{1}{V_{\Delta}}+\epsilon < \frac{1}{V_{3+}}$. Of course it occurs always for TmSe whatever is the valence} 
\label{differentcompounds}
\end{center}
\end{figure}

\paragraph{} To caracterize the intermediate valent state, a key parameter is the occupation number \nf\ of the trivalent configuration linked to the valence $v$ by $v=2+$ \nf\ when the valence fluctuation occurs between the divalent and the trivalent state (case of Sm, Tm and Yb) or $v=4-$ \nf\ when it happens between the trivalent and tetravalent state (case of Ce). The important difference between Sm, Tm or Yb compounds is that \nf\ can vary from 0 to 1 while in Ce intermetallic compounds : \nf\ $>0.8$ and at least long range magnetic ordering (M) occurs only for \nf\ $>0.9$\cite{Malterre}. TmSe\cite{Launois,Haen,Wertheim,Brewer} as well as SmS\cite{Wachter} and SmB$_6$\cite{Beaurepaire} in their low pressure intermediate valent gold phase have a valence near 2.6-2.7. Their trivalent limit will be reached smoothly only at very high pressure above 10~GPa for TmSe and 20~GPa for SmS and SmB$_6$\cite{Roeler,Dallera,Ogita} . As will be discussed, the striking point is that for these three systems, the change from insulating to metallic conduction at low temperature occurs when \nf $\sim 0.8$.
\paragraph{} The valence mixing between the divalent (2+) and trivalent (3+) configurations of the rare earth (RE) ions is associated to the release of an itinerant 5d electron according to the relation RE$^{2+} \Longleftrightarrow$ RE$^{3+}  + e^{-} 5d$. Experimentally, the effect of pressure is to broaden the bands and move this equilibrium to the right (increasing \nf\ ). Of course, band structure calculations are necessary to describe the real situation, but the chemical equilibrium is worthwhile to consider. In the divalent black (B) phase, the ground state is a classical insulator. Through a first order transition at $V=V_{B-G}$, a valence transition occurs to an intermediate valence (IV) gold (G) phase which is still insulating. However, under pressure the insulating gap will close for a fixed volume $V_\Delta$. At $V=V_\Delta$, metallic conduction appears for \nf $\sim 0.8$ at a volume quite larger than the volume $V_{3+}$ calculated for a pure trivalent configuration. Figure~\ref{differentcompounds} represents the location of the different compounds at ambiant pressure.

\paragraph{} In Sm compounds, the intermediate valent state occurs between a non magnetic $4f^{6}$ configuration of Sm$^{2+}$, with a zero angular momentum J and the Kramer's configuration (J=$\frac{5}{2}$) of Sm$^{3+}$ ($4f^{5}$). It looks worthwhile to predict that, as in cerium HFC, magnetic ordering will occur when the occupancy n$_f$ of the trivalent configuration approaches one. In this case, following the Doniach model (see \cite{revuedeJacques}), the Kondo coupling should be small enough, so that the Kondo energy becomes smaller than the RKKY energy. However, recently it was shown by use of a microscopic hyperfine technique as nuclear forward scattering and a macroscopic probe as ac microcalorimetry that  magnetic ordering  occurs already for a rather large departure from \nf $=1$\cite{BarlaSmS, Haga, BarlaSmB6}. Up to \nf $\leq 0.8$, the 4f wavefunction seems to be renormalized to the 2+ configuration while above \nf $\sim 0.8$, it seems linked to the 3+ configuration. Furthermore, this is related to the conduction properties : insulating below \nf $\sim 0.8$ and metallic above. 

\paragraph{} In Tm chalcogenides, the ground state of the divalent configuration (\nf\ $=0$, case of TmTe)  is insulating, and becomes metallic for the trivalent form. In the IVC (case of TmSe) with low \nf\ (\nf\ $\leq 0.8$), the many body effects of the correlation lead to the survival of an insulator. In the specific case of Tm, the novelty is that mixing occurs between two configurations with non zero angular momentum. The divalent one (\tmd\, $4f^{13}$) is a Kramer's configuration with $J=\frac{7}{2}$ which leads to a doublet or a quartet crystal field ground state, while the trivalent one (\tmt\, $4f^{12}$) is a non Kramer's ion with $J=6$ which may lead to a singlet crystal field ground state. The pressure induced collapse of the insulating state is associated with a change in the magnetic structure at $p_{\Delta}\sim 3~$GPa\cite{Ribault}. Below $p_{\Delta}$, i.e. for \nf $\leq 0.8$, the ground state is insulating, like in the low pressure intermediate valence phase of SmS and SmB$_6$, and antiferromagnetic of type I with properties basically given by a dressing towards a divalent renormalization (insulating conduction, doublet degeneracy of the local magnetic level). Above $p_{\Delta}$, the ground state is metallic (like TmS, or SmS and SmB$_6$ at high pressure), again antiferromagnetic, but of type II, with properties renormalized to the trivalent configuration. A surprising report was that near p$\sim 6$~GPa, TmSe may become insulating again\cite{Ohashi1,Ohashi2}. 

\paragraph{} In this paper, we present a detailed study of the high pressure phase diagrams of TmSe, TmS and \smb\ . Since for the two first cases, specific heat is already well known for $p=0$, and also interplay occurs between pressure and ligand effects, those compounds allow to verify the faisability and difficulties of high pressure microcalorimetry experiments. As TmSe is already magnetically ordered at ambiant pressure, up to 3~GPa, one may expect a signal in the ac calorimetry equivalent to ambiant pressure. Above 3~GPa, the signal may change as the signal may be normalized to the 3+ configuration like in TmS. Special attention is given on the pressure range around 6~GPa. The evolution of $T_N(p)$ of TmSe above 3~GPa will be compared to the nearly trivalent TmS. In \smb\ , we found evidence for a magnetically ordered ground state for $p>8$~GPa. However, a homogeneous ground state appears only above 10~GPa.

\paragraph{} The paper is organized as follows. First we will discuss details of the ac calorimetry technique. Then, the experimental results on TmSe, TmS ans \smb\ will be presented and an experimental conclusion will be given. In the last part, the influence of the valence on the appearance of magnetic order will be discussed in detail and a comparison to the well known high pressure phase diagrams of Ce and Yb Kondo lattice will be given.

\section{Experimental}

\begin{figure}
\begin{center}
\includegraphics[width=5cm]{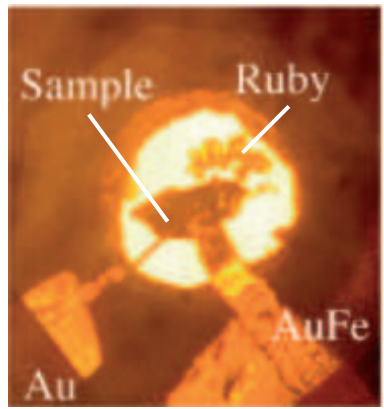}
\caption{Zoom on the high pressure cell. A thermocouple made of Au and AuFe is welded on the sample. Argon is used as pressure medium. The pressure is measured due to the fluorescence shift of ruby. The diamater of the hole is about 350~$\mu m$.}
\label{photo}
\end{center}
\end{figure}

\paragraph{}  The TmSe and TmS single crystals were  prepared by F. Holtzberg in IBM research center, New York, and  samples of the same batch have been intensively studied previously in CNRS Grenoble\cite{TmSeTmSauCNRS}. \smb\ single crystals were grown in CEA Grenoble, out of an aluminium flux. The samples studied were cleaved to be approximately 200*100*50$\mu m^3$ in size. The high pressure experiment were performed in a diamond anvil pressure cell (see figure~\ref{photo}). Argon is used as a pressure transmitter. The pressure is measured at low temperature by the shift of the ruby fluorescence line. In the ac calorimetry, a laser is used as heater. The beam is modulated using a mechanical chopper which works in the frequency range 50~Hz$<f<$5000~Hz. The temperature oscillations of the sample are measured with a Au/AuFe(0.07\%) thermocouple which is spot welded on the sample. In the case of TmSe, it was glued with very diluted General Electric varnish. It is important that the thermocouple is welded in one point to avoid contributions of the thermoelectric power of the sample itself. A lock-in amplifier is used to measure the voltage of the thermocouple.

The measurements were performed in a $^4$He bath cryostat. 

\paragraph{} This experimental situation can be described by a first order model neglecting all internal time constants between sample, heater and thermometer\cite{accalorimetry} : $T_{\rm ac} = \frac{P_0}{\kappa + i \omega C}$, where $T_{\rm ac}$ is the amplitude of the temperature oscillation, $P_0$ the average  power transmited, $\kappa$ the thermal conductivity to the bath and $C$ the specific heat. Even if the leak $\kappa$ is unknown, the phase measured by the lock-in is supposed to give the possibility to extract the value of the specific heat. $C=\frac{P_0.S}{V.\omega} \sin (\phi -\phi_0)$, where $V$ is the voltage of the thermocouple, $S$ its relative thermopower and $(\phi - \phi_0)$ the phase of the signal. If we want to minimize the importance of the phase correction, the choice of the frequency is crucial, as it balances the importance of the specific heat compared to the leak in the signal measured.  From this point of view (without considering noise problems due to a decrease of the signal at high frequency), the frequency should be the highest possible. However, the experiment will show that this model is no longer valid at higher frequencies. If the frequency is too high, the sample decouples from the thermocouple and, the thermocouple can be directly excited by the laser and measure only its own temperature at high frequency\cite{chrisetmam}. 
\begin{figure}
\begin{center}
\includegraphics[width=7cm,clip]{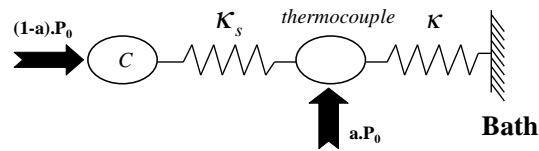}
\caption{Schematic view of the thermal system in the pressure cell. The laser gives the power $aP_0$ and $(1-a)P_0$ respectively to the thermocouple and the sample}  
\label{schema}
\end{center}
\end{figure}
The next step is to include a thermal conductivity $\kappa_S$ between the sample and the thermocouple and to consider that a small proportion $a$ of the power is directly received on the thermocouple. In this situation (see figure~\ref{schema}), $T_{\rm ac}$ can be reestimated:\cite{TheseMAM} $T_{\rm ac} = \frac{P_0.(1-\frac{\kappa_{eff}}{\kappa_S})}{\kappa_{eff}}\frac{1 + i \omega a \frac{C}{\kappa_S}}{1 + i \omega \frac{C}{\kappa_{eff}}}$ with $\kappa_{eff}=\frac{\kappa \kappa_{S}}{\kappa+\kappa_S}$ representing the total parallel thermal conductivity of leak. 
Three different limits can be distinguished : 
\begin{itemize}
\item At low frequency, if $\omega C \ll \kappa_{eff}$ then $T_{\rm ac}= \frac{P_0.(1-\frac{\kappa_{eff}}{\kappa_S})}{\kappa_{eff}}$. The value of the basic model is recovered : the phase of the signal is nearly zero and the inverse of the module is small.
\item For the intermediate regime $\kappa_{eff} \ll \omega C \ll \kappa_S$ we recover also the basic model $T_{\rm ac}= \frac{P_0.(1-\frac{\kappa_{eff}}{\kappa_S})}{\kappa_{eff}+i\omega C}$. In good conditions, if frequency becomes high enough compared to the leak, the phase reaches nearly -$\frac{\Pi}{2}$.
\item Finally, at high frequency, for $\kappa_{S} \ll \omega C$, $T_{\rm ac}=(1-\frac{\kappa_{eff}}{\kappa_S})\frac{a P_0}{\kappa_S}$.The phase reaches zero and the module decreases again. Physically, the thermocouple is decoupled from the sample.
\end{itemize}

\begin{figure}
\begin{center}
\includegraphics[width=7cm]{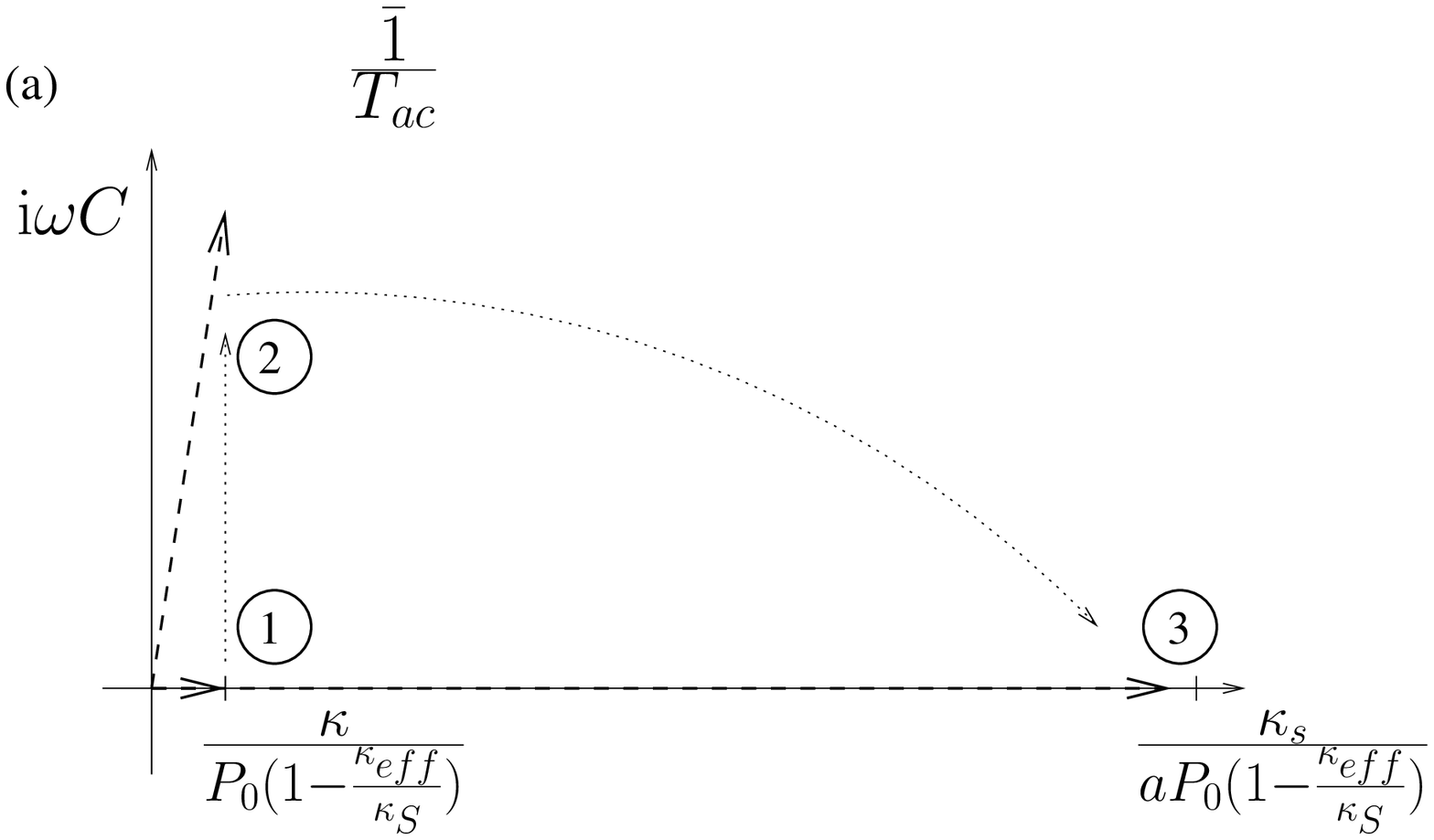}
\includegraphics[width=7cm]{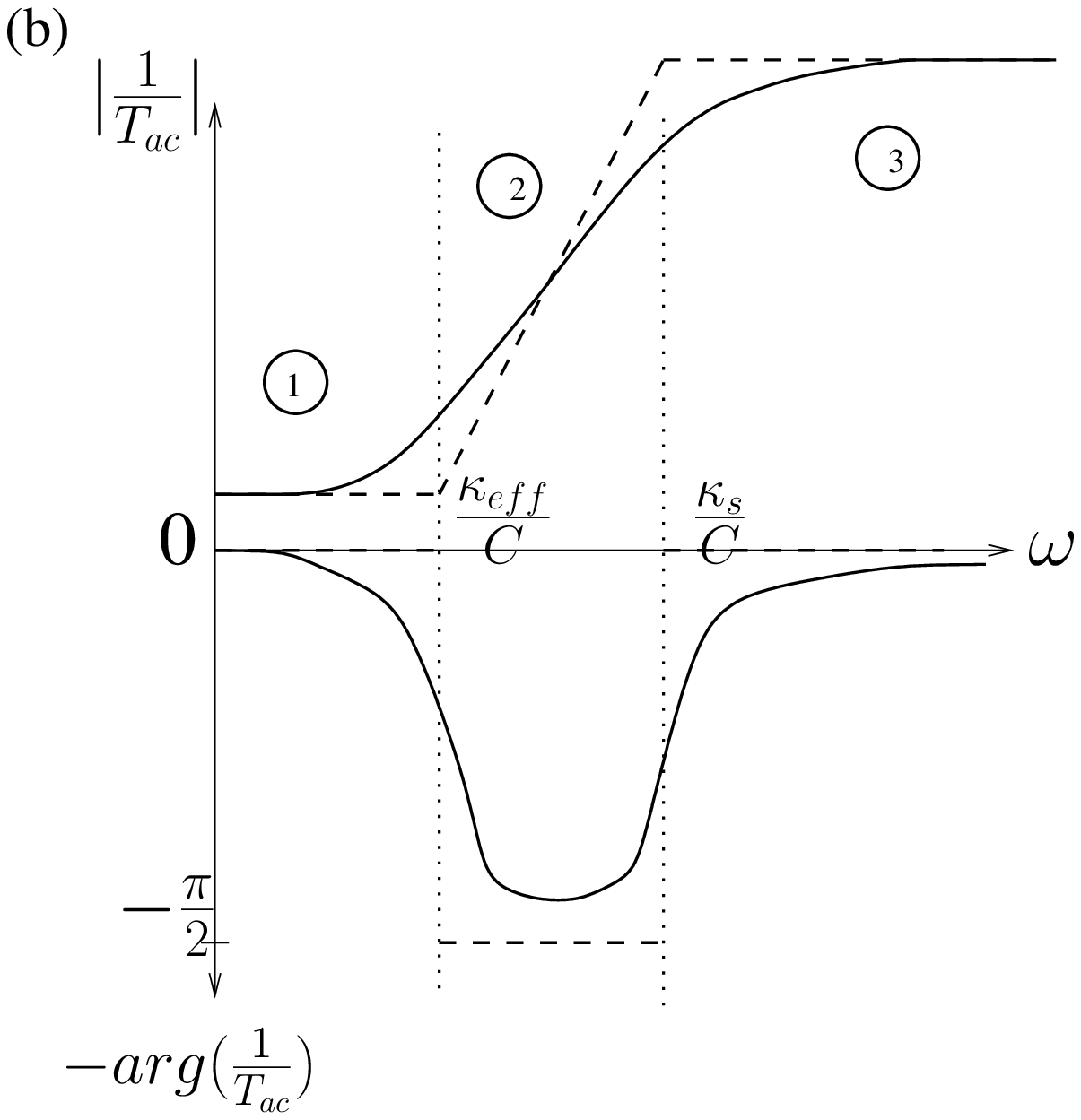}
\caption{Part (a) shows graphically the inverse of $T_{\rm ac}$ in the complex representation. Leak phenomena is on the real axis whereas capacitive effects are on the imaginary axis. The three limit cases are : (1), capacitive effect is negligible and the power $P_0$ is transmited to the bath with the leak $\kappa$ ; (2), capacitive effect becomes dominant and the component $\omega C$ is added ; (3) The sample is decoupled from the thermocouple. the power received is only the fraction $a P_0$ and the main leak is still $\kappa_S$, towards the sample. In part (b), the schematic shape of the phase and the module of $\frac{1}{T_{\rm ac}}$ is deducted from the evolution drawn in part (a). The vertical dashed lines show the cut off frequencies $\frac{\kappa_{eff}}{C}$ and $\frac{\kappa_{S}}{C}$ which indicate the change of regime corresponding to (1), (2) and (3).}  
\label{complex}
\end{center}
\end{figure}

To view more clearly the frequency dependence of the system, let us consider the complex number $\frac{1}{T_{\rm ac}}$. The phase measured by the lock in is directly the opposite of the phase of this complex number, and the signal $\frac{1}{V}$ is directly linked to the module of this complex number. Part (a) of the figure~\ref{complex} explains the different regimes depending on the frequency. From that picture, we can roughly draw the shape of the phase and of the inverse of the module (see part (b) of the figure~\ref{complex})

 Moreover, if we consider the variable change $\omega\leftrightarrow \omega C$, the shape of the dependance in $\omega$ of the argument and module of $\frac{1}{T_{ac}}$(figure~\ref{complex}b) can be expanded to the dependance in $\omega C$. Then, considering a jump in the specific heat $C$ at the magnetic transition, the phase will be changed differently at high and low frequency. Around $\omega=\frac{\kappa_{eff}}{C}$ the signal in the phase will be a negative peak, but around $\omega=\frac{\kappa_S}{C}$, the signal of the phase can be a positive peak. This will be confirmed later by the experimental results.
   
 Thus, the best frequency for the measurement is between this two cut-off frequencies. Typically, the best frequency  was about 90~Hz for TmSe, 800~Hz for TmS and 4500~Hz for \smb. Assuming that the specific heat of TmSe is higher than that of TmS which is higher than that of \smb\  (at least at low pressure as indicated in figure~\ref{c0}), this support the model since in the conditions of measure $C \omega$ stays roughly constant. 

\paragraph{} Nevertheless, even if the behaviour of the phase is understood, the incertitude on the reference phase $\phi_0$ and the complex influence of pressure keep the situation delicate. Therefore, in the following, we will usually estimate the specific heat via the simplest expression : $C=\frac{P_0.S}{V.\omega}$. 

\paragraph{} The  main point of the apparatus is the possibility to change the pressure at low temperature and also to use a excellent hydrostatic medium (Ar or He). To improve the faisability of the difficult microcalorimetric measurement under hydrostatic pressure, the choice has been made to minimize the number of electrical leads and thus, to use a laser as heater. The advantage of the technique is to give the pressure variation of the Néel temperature with a great accuracy i.e. a large set of pressure.  If it is an excellent method to determine the phase diagram, the difficulty is to extract the specific heat in absolute units.

\section{Results} 
\subsection*{Preliminary results}
\begin{figure}
\begin{center}
\includegraphics[width=7cm]{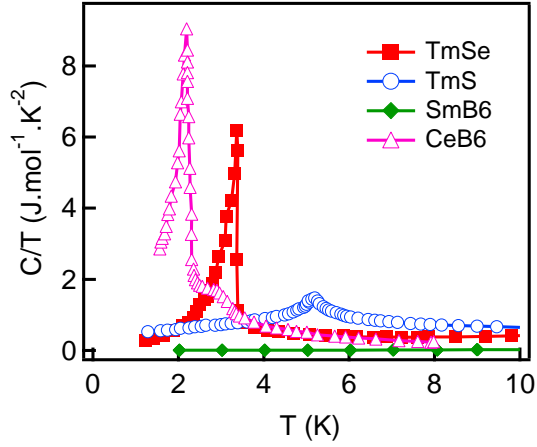}
\caption{Specific heat divided by temperature measured at ambiant pressure for TmSe\cite{Berton}, TmS\cite{TmSPzero}, SmB$_6$(measured on our sample) and CeB$_6$\cite{CeB6}}
\label{c0}
\end{center}
\end{figure} 
\paragraph{} Before discussing the specific heat under pressure, we present the specific heat at ambiant pressure for the different systems in figure~\ref{c0}. The behaviour of $C$ for the two Tm compounds is quite different. For TmSe, the specific heat has a sharp anomaly at \tn \cite{Berton}. TmS is metallic and the crystal field ground state may be a singlet. Here, large fluctuations are already oberved above \tn \cite{TmSPzero}. In the other case, as \smb\ is non magnetic we have reported here the results for CeB$_6$\cite{CeB6} in order to have an idea of the amplitude of the signal under pressure. The comparison is worthwhile as both ($4f^{1}$) Ce$^{3+}$ and ($4f^{5}$) Sm$^{3+}$ are Kramer's ions with the same angular momentum $J=\frac{5}{2}$ with a lifting of the degeneracy by the crystal field in a $\Gamma_{7}$ doublet and a $\Gamma_{8}$ quartet. The successive transitions observed for CeB$_6$ are now well understood by a cascade from paramagnetism to quadrupolar ordering at $T_{N_{1}}\sim 2.9$~K and to dipolar ordering at $T_{N_{2}}\sim 2.2$~K, the crystal field ground state being a $\Gamma_{8}$ quartet\cite{Shiina98}.
\subsection*{TmSe}

\begin{figure}
\begin{center}
\includegraphics[width=7cm]{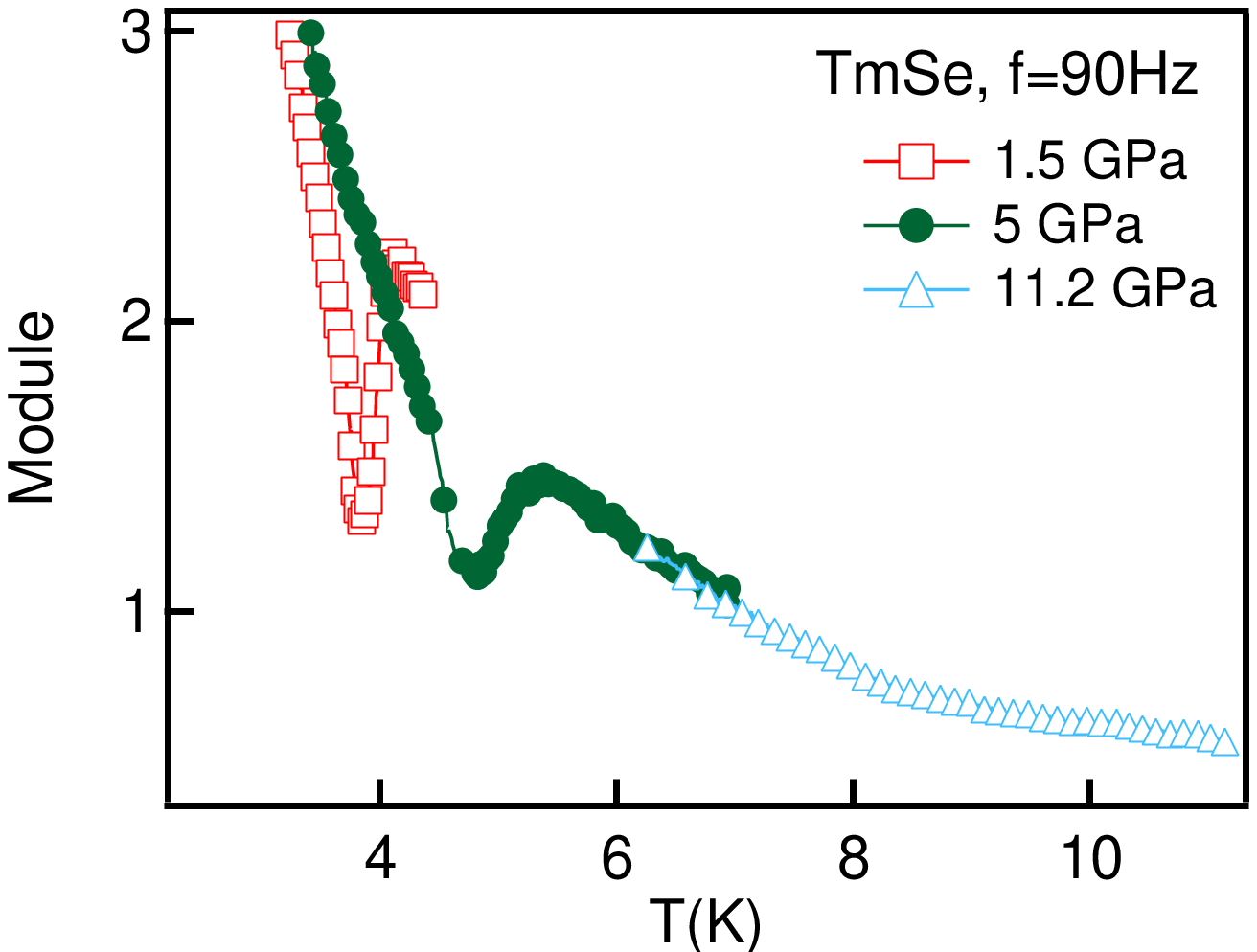}
\includegraphics[width=7cm]{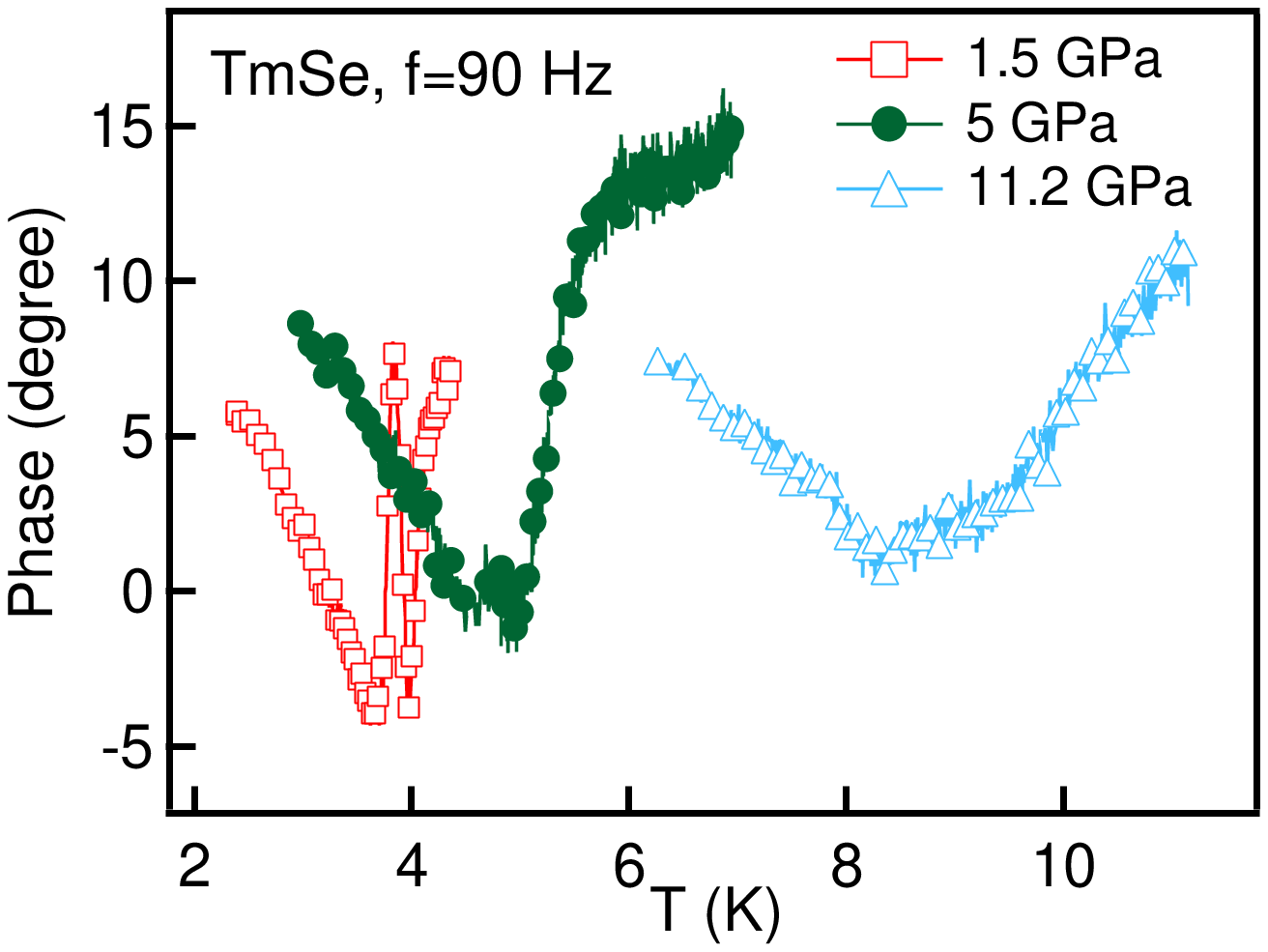}
\caption{Raw data measured for TmSe, for several pressures (1.5, 5 and 11.2~GPa.). Module data are normalized at high temperature ; therefore, the three curves looks continuous after the anomaly.}
\label{raw_TmSe}
\end{center}
\end{figure}

\begin{figure}
\begin{center}
\includegraphics[width=7cm]{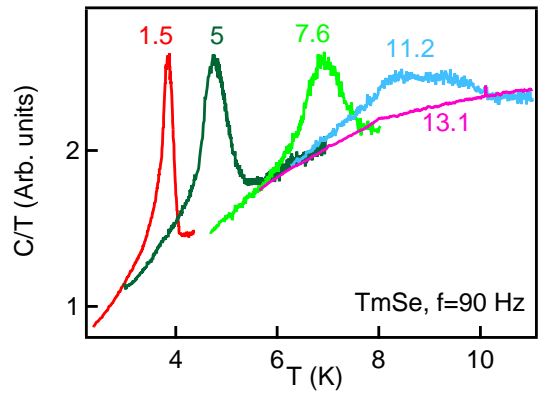}
\caption{Evolution of the specific heat anomaly of TmSe under pressure. Data are normalized at high temperature and plotted for 1.5, 5, 7.6, 11.2 and 13.1~GPa.}
\label{anomaly_TmSe}
\end{center}
\end{figure}

\begin{figure}
\begin{center}
\includegraphics[width=6cm]{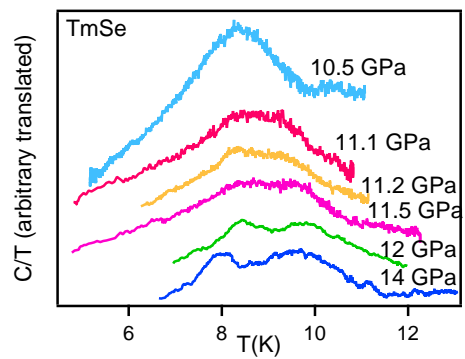}
\caption{Observation of the spliting of the specific heat anomaly of TmSe at high pressure.}
\label{split}
\end{center}
\end{figure}

\paragraph{} The temperature dependence of the specific heat of TmSe has been measured in a wide pressure range (1 to 14~GPa). Raw data are plotted in figure~\ref{raw_TmSe} for different pressures.
The modulus and also the phase show a clear anomaly at the magnetic transition. The measurement has been realised at low frequency, so that the magnetic anomaly is seen in the phase as a negative peak. A second very sharp positive peak is also observed, inside the first negative peak, especially at low pressure. A simulation\cite{TheseMAM} shows that the huge value of the specific heat jump in TmSe can induce this second positive peak changing from the regime of low $\omega C$ to the one of high $\omega C$. Nevertheless, the strongly negative peak of the phase shows that we are in the low frequency regime where phase correction is supposed to be used. Moreover, as the signal is huge on the modulus, the phase correction is not significant (This is explained because close to $-\frac{\Pi}{2}$ the sinus is not really sensitive). Thus we present here the simplest estimation of the specific heat $C=\frac{P_0.S}{V.\omega}$. Some of the calculated curves are plotted in figure~\ref{anomaly_TmSe}. 
\paragraph{} A first observation is that the magnetic anomaly is very well defined, so that, the Néel temperatures can be easily extracted. To define \tn\ we choose the maximum of the anomaly. Furthermore, we found  an unexpected broadening of the anomaly as the pressure increases. This appears below 10~GPa, when hydrostaticity is still very good (less than 0.1~GPa of variation in the cell)\cite{jean}. This broadenning cannot be explained by pressure inhommogeneities as $\frac{dT_N}{dp}$ is small. Figure~\ref{split} shows $\frac{C}{T}$ for pressures above 10~GPa. The data indicate a splitting of the magnetic anomaly, which can be followed under pressure. 

\begin{figure}
\begin{center}
\includegraphics[width=7cm]{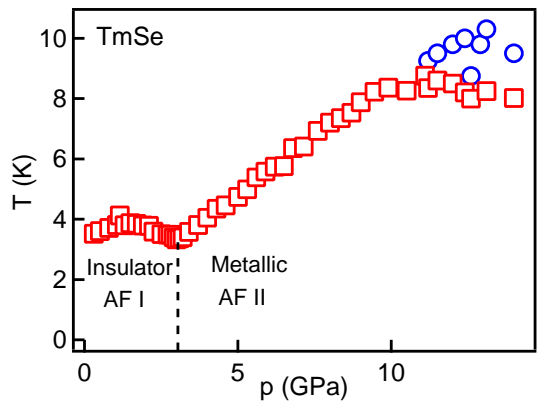}
\caption{(p, T) Phase diagram of TmSe. The squares represent the maximum of each specific heat anomaly, and the circles indicate the second maximum observed at high pressures. The dashed line show the phase transition coresponding to the slope change.}
\label{phasediagram_tmse}
\end{center}
\end{figure}

The resulting phase diagram is then represented in figure~\ref{phasediagram_tmse}
\paragraph{} The phase diagram can be distinguished in three parts. At low pressures, the evolution of T$_N$ is quite flat and a maximum can be seen around 1.3~GPa. Then, a break in the slope around 3~GPa corresponds to the pressure of transition from the insulating AF1 phase to the metallic AF2 phase. The second magnetic structure is caracterized by a linear increase of the Néel temperature with pressure.

\paragraph{} At low pressure, our data are completely consistent with previous resistivity measurements\cite{Ribault,Ohashi1,Ohashi2}. The important observation is the continuous increase of \tn\ with pressure at high pressure. Contrary to recent resistivity measurements who showed a discontinuity in the N\'eel temperature around 6~GPa\cite{Mignot}, no anomaly in $T_N(p)$ is seen in our data. Actually, our observation is consistent with a release of the 5d electrons near 3~GPa. Recent neutrons measurements\cite{Mignot} confirm this idea as no change in the magnetic structure is found at 6~GPa. 
Finally, an interesting splitting of the magnetic anomaly is observed at high pressure, above 10~GPa. The evolution of the signal shape was detailled in figure~\ref{split}. The origin of this splitting and the new phase is not clear. This observation pushs us to study TmS, which can be seen as high pressure analog of TmSe. In TmS, evidence has been reported for two different magnetic phases~\cite{OashiTmS} around 5~GPa.
\subsection*{TmS}

\begin{figure}
\begin{center}
\includegraphics[width=7cm]{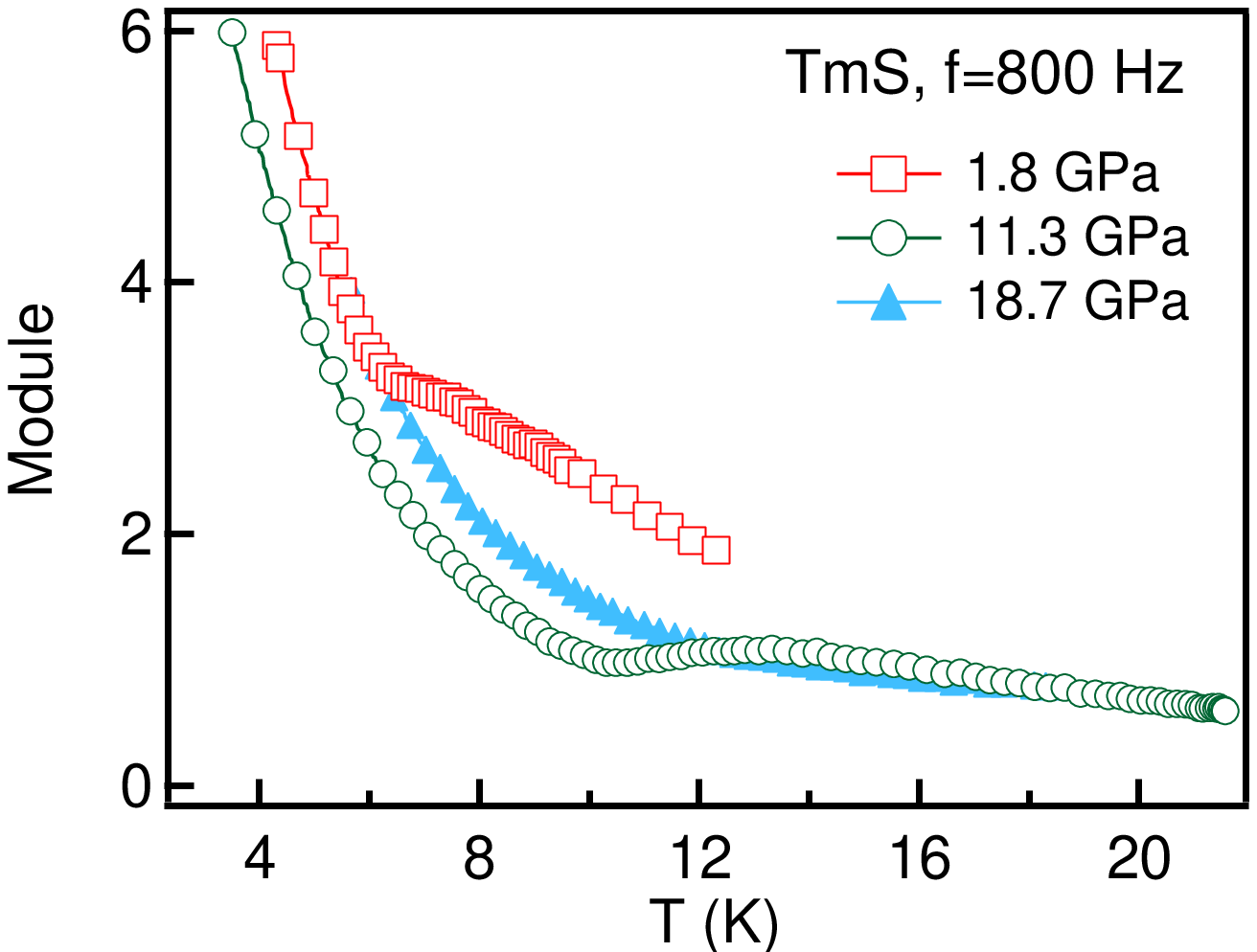}
\includegraphics[width=7cm]{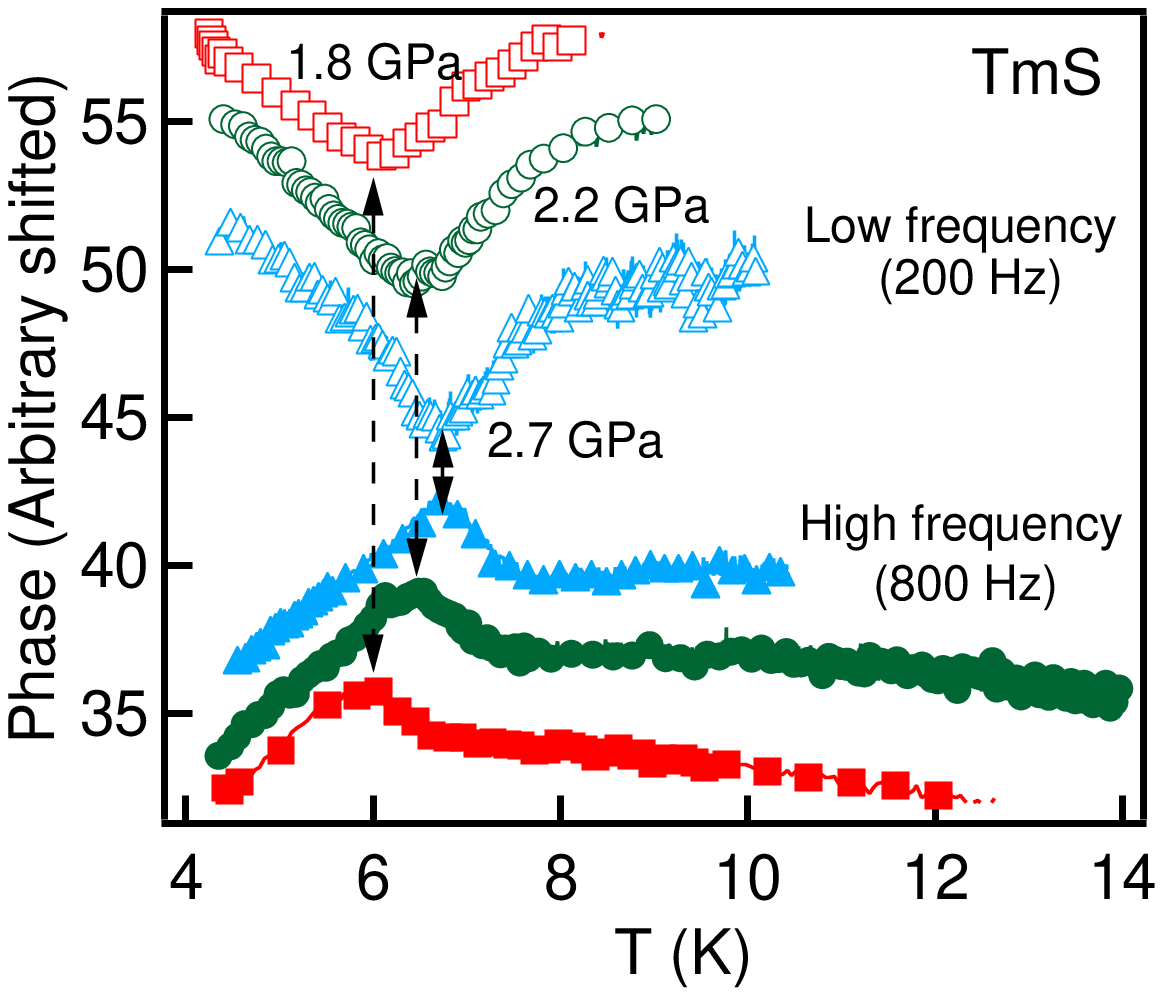}
\caption{Raw data measured for TmS. The module has been drawn for 1.8, 11.3 and 18.7~GPa. The signal has been followed until very high pressure, but at the end, it desapears as shows the curve at 18.7~GPa. Then, the behaviour of the phase has been detailled  : the measurement at low frequency ($200$~Hz) and high frequency ($800$~Hz) have been compared and followed with pressure from 1.8~GPa to 2.7~GPa. For the picture the phase have been arbitrary shifted.}
\label{raw_tms}
\end{center}
\end{figure}

\begin{figure}
\begin{center}
\includegraphics[width=7cm]{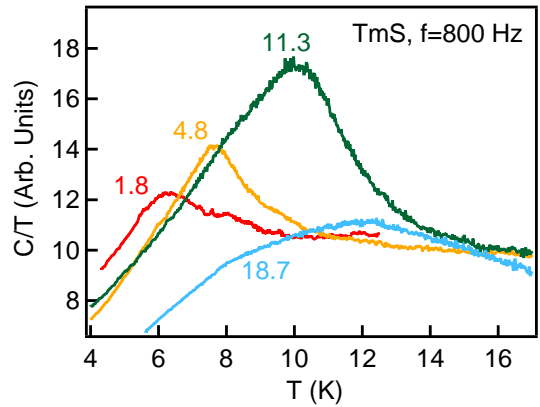}
\caption{Evolution of the magnetic anomaly of TmS under pressure ; specific heat divided by temperature has been normalized at high temperature and plotted for different pressures : 1.8, 4.8, 11.3 and 18.7~GPa}
\label{anomaly_tms}
\end{center}
\end{figure}

\paragraph{} The specific heat of TmS was measured up to 19~GPa. Raw data are plotted for several pressures in figure~\ref{raw_tms}.
The behaviour of the phase is detailled for the low pressures. The previous explanation is confirmed : we can observe two different regimes for the phase, depending if the measurement is performed at low or high frequency. This is really reproducible and stable with pressure change. That confirms that the feature occuring on the phase is very useful to detect the magnetic transition. Unfortunately, in the low frequency regime, the feature on the modulus is very small and don't allow us to extract a good shape of the specific heat. On the other hand, figure~\ref{raw_tms} shows that the module measured at high frequency is more clear. Even if the first order model is valid only at low frequency, figure~\ref{complex}b shows that the evolution of the module is still monotonous even after the first cut off. Thus, in order to avoid a correction with an arbitrary phase $\phi_0$, we prefer to show the estimation at zero order of the specific heat at 800~Hz. Some selected pressures are shown on figure~\ref{anomaly_tms}.

\begin{figure}
\begin{center}
\includegraphics[width=7cm]{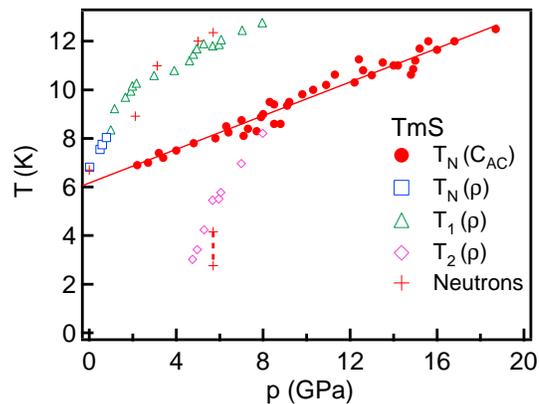}
\caption{Phase diagram of the compound TmS ; The maximum of the magnetic anomaly measured under pressure has been plotted (full circles) in the same time as previous neutron scattering\cite{Mignot}(crosses) and resistivity measurements\cite{OashiTmS} (empty symbols). }
\label{phasediagramm_TmS}
\end{center}
\end{figure}

\paragraph{} Increasing the pressure,  the maximum is shifted to higher temperature, from 6 to 12~K. Until 15~GPa the signal is only slighty broadened, and still very clear, but at higher pressure, the signal decreases. The phase diagram of TmS is shown in figure~\ref{phasediagramm_TmS}. Anomalies found in previous resistivity measurements\cite{OashiTmS} and neutron scattering\cite{Mignot} have also been plotted.  $T_1$ and $T_2$ are kinks observed in the resistivity curve. $T_1$ looks linked to $T_N$ and $T_2$ indicates a new phase which has also been evidenced by neutron scattering.
\begin{figure}
\begin{center}
\includegraphics[width=7cm]{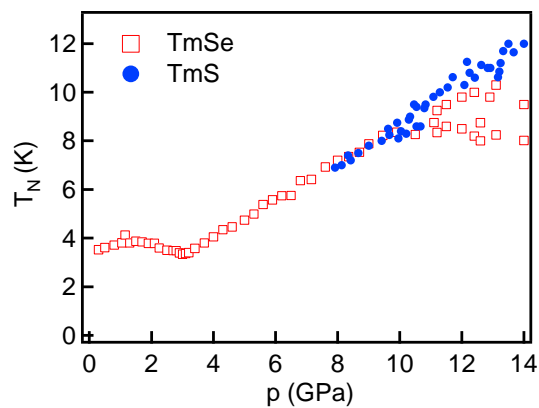}
\caption{Combination of the phase diagram of both TmSe and TmS. For the abscisse axis, we have choosen a typical volume linked to the pressure in the TmSe compound. That means that pressure for TmS has been renormalized .}
\label{phasediagramm_combine}
\end{center}
\end{figure}
Our study indicates a linear $p$ increase of the N\'eel temperature. This observation differs from published results obtained by resistivity or neutron scattering experiments. The sensitivity of TmS to defects is well known. At ambiant pressure, the value of \tn\ is sample dependent and varies between 5.2~K and 7.05~K\cite{TmSdefect,TmSPzero}. Our sample comes from the same batch than the crystal measured in reference\cite{TmSPzero} where excellent agreements was found between different methods in the \tn\ determination. 
The second anomaly below \tn\ observed by neutron scattering in the $p$ range above 5~GPa is due to a "lock-in" transition from an incommensurate to a commensurate structure. Therefore, if entropy is just slighty changed, it might be not detected by our specific heat measurement. Of course, an open question is again here the reproductibility of this second anomaly   with the defects' content

 In order to compare these results to  TmSe, we have scaled the pressure applied on TmS, into an equivalent pressure applied on TmSe, to obtain the same volume. The pressure range has been shifted of 7~GPa, corresponding to the value where TmSe is more or less trivalent, and then normalized by the ratio of the compressibility of the two compounds (1.5 $10^{-6}$~bar$^{-1}$ for TmS and 3.5 $10^{-6}$~bar$^{-1}$ for TmSe from reference\cite{compress1,compress2}). The resulting phase diagram is plotted in figure~{\ref{phasediagramm_combine}}. 

\tn\ of TmS scales very well to TmSe. Of course, the points of TmSe don't follow completely the same alignement at too high pressure: the TmSe measurements themselfs have to be renormalized at very high pressure as the compressibility of TmSe decreases\cite{compress1,compress2}.

\subsection*{SmB$_6$}

\begin{figure}
\begin{center}
\includegraphics[width=7cm]{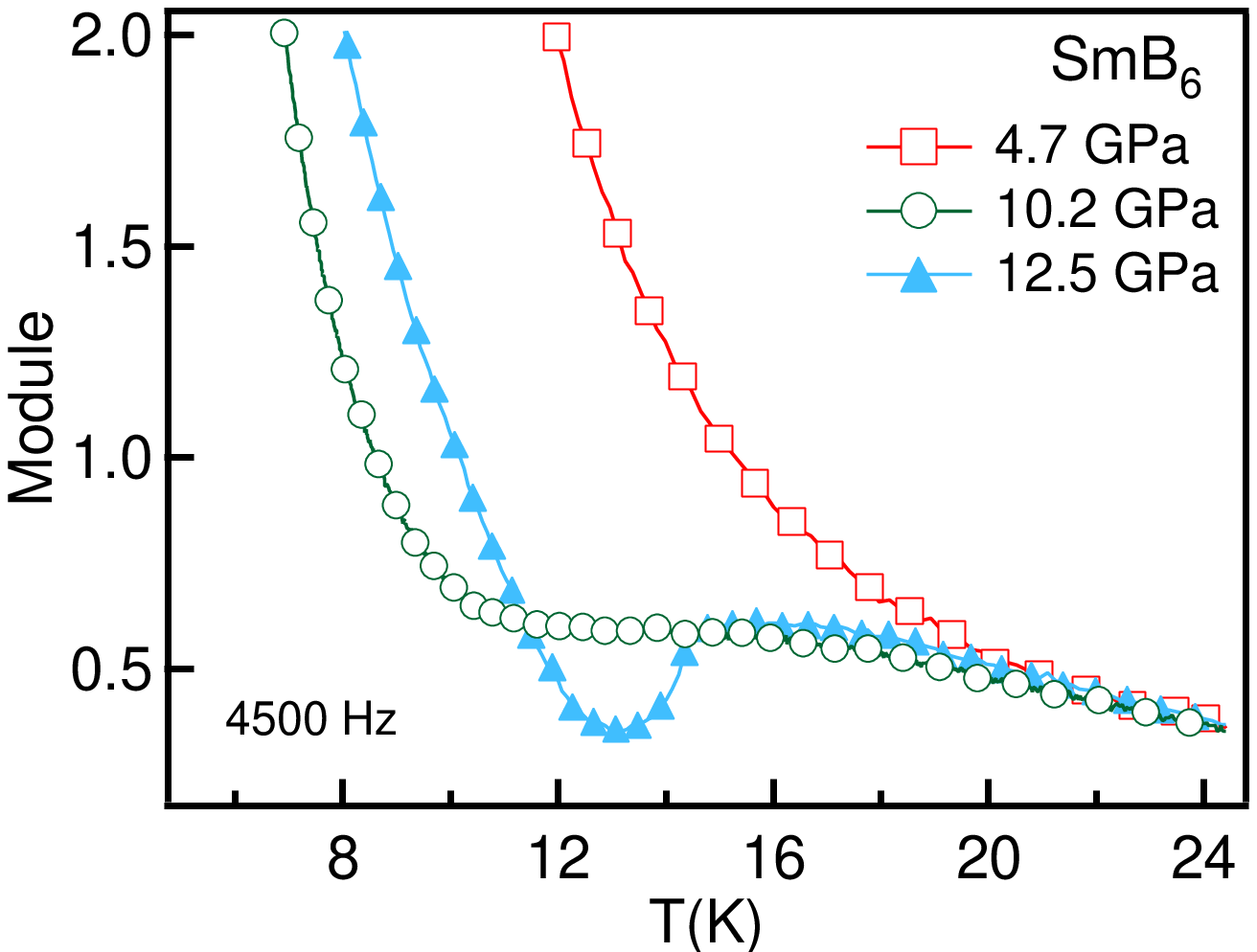}
\includegraphics[width=7cm]{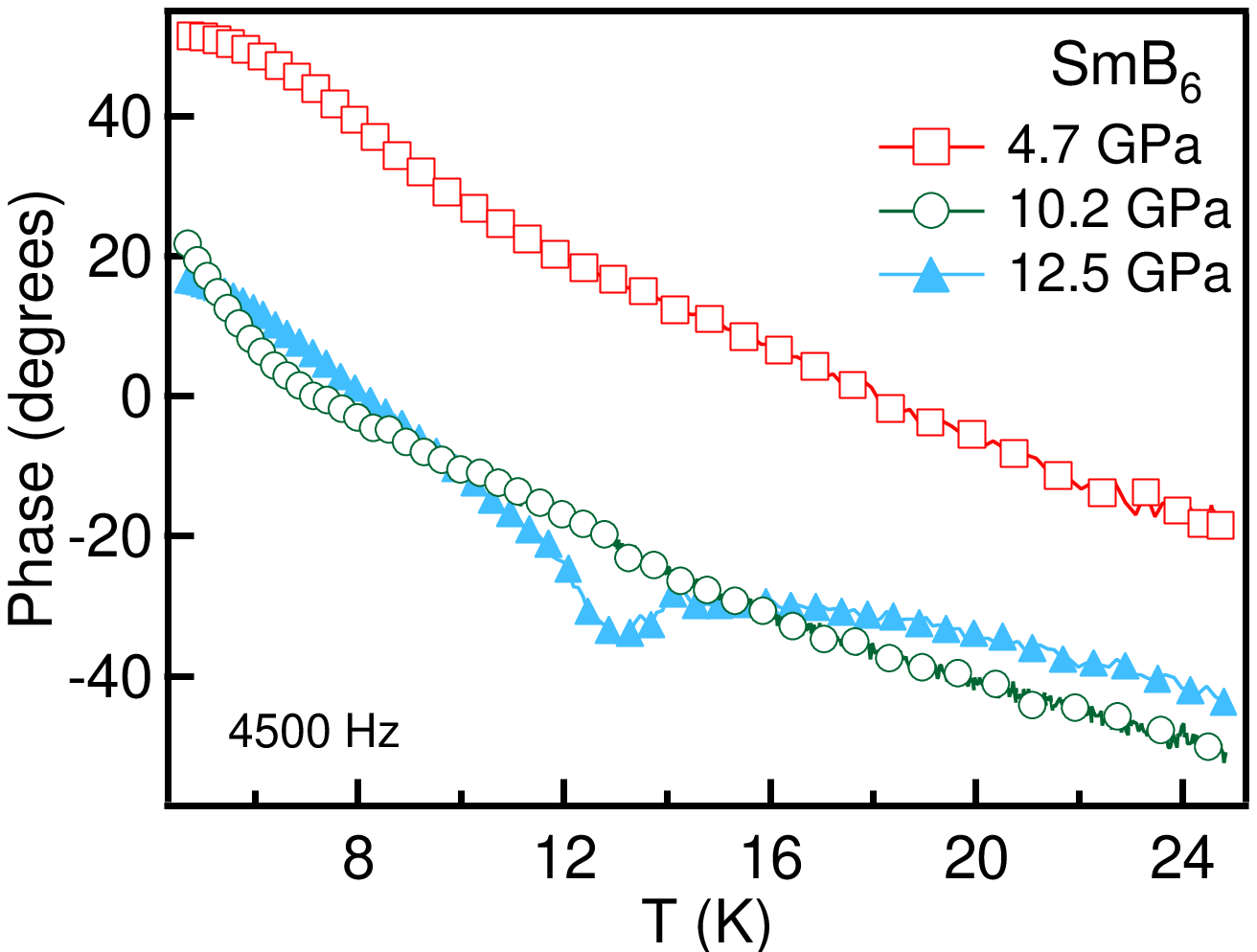}
\caption{Raw data measured for SmB$_6$ for several pressures (4.7, 10.2 and 12.5~GPa.). Module data are normalized at high temperature}
\label{raw_smb6_2}
\end{center}
\end{figure}

\begin{figure}
\begin{center}
\includegraphics[width=7cm]{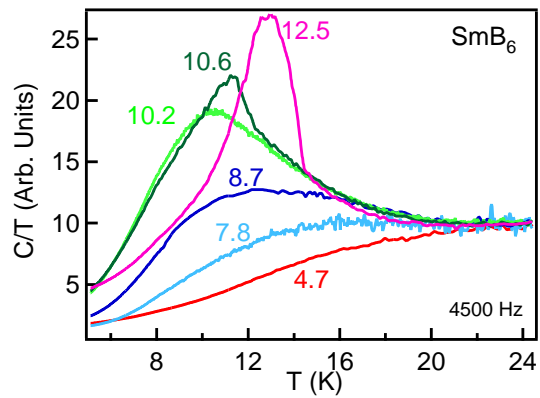}
\caption{Growth of the magnetic anomaly of \smb\ under pressure ; specific heat divided by temperature has been normalized at high temperature and plotted for different pressures : 4.7, 7.8, 8.7, 10.2, 10.6 and 12.5~GPa}
\label{anomaly_smb6}
\end{center}
\end{figure}

\begin{figure}
\begin{center}
\includegraphics[width=7cm]{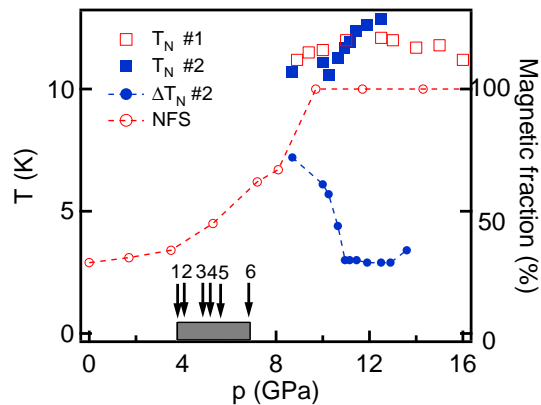}
\caption{Phase diagram of \smb. The N\'eel temperature (dark square)  and the broadening $\Delta$T$_N$ have been ploted in Kelvin (the broadening is the width ot the anomaly peak at half of the height). We have also plotted \tn\ for another cell measured previously (light square). These results are compared to the magnetic fraction measured by NFS \cite{BarlaSmB6}. The vanishing of the gap is also represented by arrows coresponding to different studies : 1-Sample given by K. Flachbart measured in the laboratory, 2-ref.\cite{Gabani}, 3-ref.\cite{Moshchakov}, 4-ref.\cite{Cooley}, 5-Sample grown by G. Lapertot and measured in the laboratory and 6-ref.\cite{Beille}. The dashed box shows the wide pressure range corresponding to the collapse of the hybridisation gap observed in different samples}
\label{smb6diagram}
\end{center}
\end{figure}

\paragraph{} Finally, similar experiments have been performed for SmB$_6$. Long range magnetic ordering has been found above 8~GPa\cite{BarlaSmB6}. The features observed on the raw data are already clear. They have been plotted in figure~\ref{raw_smb6_2} 
and the magnetic anomaly shows clearly up in the modulus. The feature in the modulus is so huge, and we never reach the "high frequency regime" with inversion of the phase, even for the highest frequency allowed by the set up. Thermal contact between the sample and the thermocouple was very good. Therefore, the specific heat has been extracted only from the modulus measured at very high frequency, and a selection of the results have been plotted in figure~\ref{anomaly_smb6}. With increasing pressure, the anomaly gets more and more pronounced. Contrary to the case of TmSe, the peak gets sharper, even above 10~GPa. 
\paragraph{} The phase diagram of \smb\ is shown in figure~\ref{smb6diagram}. We choosed as criterium for \tn\ the maximum of the anomaly in $\frac{C}{T}$. In order to look more carefully at the change of the shape of the signal, we have also investigated the broadening of the anomaly which is plotted in figure~\ref{smb6diagram} too.
   
\paragraph{} The evolution of the broadening, shows that the anomaly peak is first very broad and then sharper. Moreover, a change of regime appears around 10~GPa. This change is significant as we can observe a clear change in the slope of the broadening i.e. roughly at the pressure where 100\% of magnetic sites has been detected by NFS\cite{BarlaSmB6}.

\subsection*{Experimental conclusion}

\paragraph{} It has been shown that the experimental set up of the cell is critical to obtain correct shapes of the specific heat. Especially the link between the thermocouple and the sample must be very good. The main incertitude concerns the knwoledge of the reference phase $\phi_0$. With that information, it could be possible to correct the variation due to the leak but non monotonous behaviour of the phase before 4~K (certainly due to a $T$ dependence of $\kappa(T)$) has discouraged us to associate $\phi_0$ with the phase measured at low temperature. So that, the extraction of an absolute value of the specific heat remains difficult and as the phase correction is generally small, we have prefered to show here estimation derived only from the module. Nevertheless, the method is very useful to detect the pressure induced phase transitions (here, long range magnetism), and the in situ pressure generation gives a fine pressure tuning. Thus this technique is well adapted to draw phase diagrams. 


\paragraph{} The main experimental problem is to understand the broadening and the loss of the magnetic anomaly under pressure. One could imagine that the thermal contact between the sample and the thermocouple is one of the issue. But, as the systems were well welded, we don't believe in a loss of the contact. Another consideration is the behaviour of the thermal leak, as it can become huge at high pressure. The first guilty phenomenum is the argon conductivity, if we extrapole some conductivity measurements done at higher temperature\cite{argon}, the conductivity increases with more than a factor 10 between 1 and 10~GPa. At 1~GPa, the two terms $\omega C$ and $\kappa$ can already be estimated of the same order (10$^{-3}$~W K$^{-1}$), so that a factor 10 will be a huge effect for the relative signal measured at 10~GPa. Of course this doesn't explain the relative sudden character of the effect as $\kappa$ increases roughly linearly. If we consider the big compressibility of the argon\cite{argonbis} we can expect a reduction of the volume of the pressure chamber of the order of 30\% .In this case, if you consider the geometry of the chamber (see figure~\ref{photo}), a possible contact could occur between the sample and the gasket at high pressure ; this could imply a big thermal contact, and a sudden increase of the thermal leak. In the case of TmSe, the anomaly is lost quickly (before 10~GPa) but for other compounds, the set up allows us to follow correctly the magnetic anomaly until around 15~GPa.

\paragraph{} For \smb\ , the situation is completely different as the broadening occurs at low pressure. There are two possible explanations for the broadening of the magnetic anomaly. First, if we assume a very sharp transition (as it seems to be, since T$_N$, nearly jumps from zero to its maximum value), the broadening could be the effect of the pressure inhomogeneity, as even a small pressure gradient  would imply a large average of the N\'eel temperatures. Nevertheless, to explain the experiment, one has to assume a quite big inhomogeneity of the order of 1~GPa. Typical deviation is about only 0.1~GPa\cite{jean}. Therefore, a sound explanation  is to consider the observed broadening as the signature of an intrinsic phenomena which may be a  mixed state linked to a first order transition. The system becomes homogeneous and reach a full long range magnetic ordering only at high pressure. This idea is consistent with NFS measurements which evidences a coexistence of two phases between 5 and 10~GPa.

\section{Discussion} 

\paragraph{} There are different approaches for the description of magnetism of TmSe and SmB$_6$ ; but, in order to make a comparison between Tm and Sm, and even with the case of Ce and Yb, we will assume that each integer valent configuration is associated to a Kondo lattice temperature \tkl\, and that it is the comparison of this characteristic energy with other energy scales like the crystal field splitting or the magnetic intersite interaction which will be led to the renormalization towards a given configuration. 

\paragraph{} The basic idea\cite{Flouquet} is that, compared to a single impurity, due to the release of an itinerant electron related to the valence mixing, a feedback occurs between the Kondo effect and the number of itinerant electrons. In analogy, to the theoretical results known for the Kondo effect of the cerium ion in the $\frac{1}{N_f}$ expansion\cite{Hewson}, we will assume that for the $3+$ configuration, $T_{K}^{3+}= (1-n_f)N_f\Delta_0$, where \nf\ is the occupation number of the trivalent state, $N_f$ the degeneracy ($N_f=2J+1$) and $\Delta_0$ the width of the virtual 4f level in the Anderson lattice related to the density of states of the light conduction electron ($N(E_f)$) and to the hybridization mixing potential ($V_{df}$) : $\Delta_0=\pi V_{df}^2 N(E_f)$. Of course $\Delta_0$ must be very sensitive to the spatial extension of the 4f orbits. One can note that the usual Kondo formula of the susceptibility  $\chi$ will be recovered for the Cerium case as it will correspond to $\chi^{3+}$ \nf \ i.e. to \tk\ $=\frac{T_{K}^{3+}}{n_f}$.

\paragraph{} In the so called \fu-\fd model (instead of the \fz-\fu model suitable for the Cerium electron case, and for the Kondo hole analog ytterbium), there are theoretical studies on TmSe\cite{Flouquet,Newns, Read,Nunes,Yafet,Saso}, for Tm impurity, with \nf\ going from zero (\fu) to one (\fd). Basically, the large $\frac{1}{N_f}$ theory leads to very similar physics than that of the \fz-\fu model with however, a maxima of the Kondo temperature around \nf $\sim 1.7$. A discussion on the Kondo effect on Sm ions can be found in reference\cite{Sakai}. For the cerium case, \tk\ will continuously increase as \nf\ decreases. Yb HFC are often viewed as the hole analog (4f$^{13}$ configuration for Yb$^{3+}$) of the Ce HFC with a decrease of \tk\ under pressure. The Tm compounds are always magnetically ordered as the exchange energy always exceeds $T_K$ , either of \tmd \ or of \tmt.

\paragraph{} Our physical picture stresses out the role of the valence mixing and the release of the 5d electron. This is the key point concerning the magnetic ground state but also the electronic ground state. That pushes us to extend the Kondo temperature formula to the lattice where the virtual bound width $\Delta_0$ is now directly related to the bare bandwidth $D$ of the 5d light conduction electrons : $\Delta_0=\alpha D$, with $D$ depending on \nf\ and $\alpha$ typically of the order of $10^{-2}$ in order to recover a narrow virtual bound state for the impurity. 
The change of the numbers of carrier will give here $D(n_f)=D_0 n_f^{\frac{2}{3}}$.

\paragraph{} If we apply this rule to TmSe, the Kondo temperature $T_{KL}$ of the trivalent and divalent configuration in a lattice will be :
\begin{itemize}
\item \tkt $=\alpha D_0 (1-$\nf $)$\nf $^{\frac{2}{3}}N_f^{3+}$ 
\item \tkd $=\alpha D_0 $\nf $^{\frac{5}{3}}N_f^{2+}$
\end{itemize}
where the degeneracy $N_f^{3+}$ and $N_f^{2+}$ are respectevely 13 and 8. 
\begin{figure}
\begin{center}
\includegraphics[width=7cm]{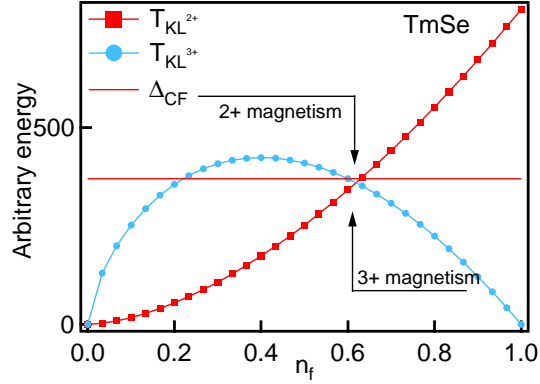}
\caption{Comparison of the Kondo lattice temperature and the crystal field spliting in arbitrary units, in the case of TmSe, for both the 2+ and the 3+ configuration. With the criteria choosen, long range magnetism is allowed if T$_{KL}$ is smaller than \cf.}
\label{simultmse}
\end{center}
\end{figure}    
Figure~\ref{simultmse} represents the Kondo temperature for the two configurations. A typical value of the overall crystal field splitting $\Delta_{CF}$ has been added to the plot. Of course, a crucial point has been to choose the ratio between \cf\ and $D_0$  to compare \cf\ with $T_{KL}$. In order to have a coherent behaviour, we put $\frac{\alpha D_0}{\Delta_{CF}}\sim 4$ which correspond to a very small effective bandwidth. Anyway, if we assume that \cf\ $\sim 100$~K\cite{Berton}, $\alpha D_0$ can be nearly the order of magnitude of 400~K. The different energies has been traced versus \nf, varying in the same way as the pressure. If there is an extra effect as a electron gap, a simple way would be to add an extra pressure dependence on $D$ ($D=0$ for $p<p_{\Delta}$).
\paragraph{} Extrapolating from the numerous studies performed on Ce HFC, the occurence of long range magnetism requires at least the recovery of usual rare earth properties, notably the full reaction to the crystal field splitting i. e.  k$_BT_{KL}<\Delta_{CF}$. Of course a main  consideration is the relative strength of the intersite exchange interaction and \tkl\ as discussed for the usual Doniach model. Long range magnetism will occur only if the energy scale of the coupling is stronger than the Kondo energy. Nevertheless, in our simple view, we compare only $\Delta_{CF}$ and \tkl . Therefore, we only indicate when long range magnetism will be possible. For each configuration,   long range magnetism will be possible while $T_{KL}$ is smaller than \cf ; that means we assume the coupling is already strong enough. Of course, the position of the intersections are very sensitive to the ratio $\frac{\Delta_{CF}}{\alpha D_0}$. Nevertheless, this basic model explains qualitatively the general shape of the phase diagram. At low pressure, \nf\ is small, and long range magnetism is due to Tm$^{2+}$ ; then at higher pressure, when \nf\ increases, this long range magnetism disapears and long range magnetism due to Tm$^{3+}$ appears. The change of regime observed at 3~GPa is well reproduced. At this critical pressure,  a critical value of \nf\ is reached, where the renormalization of the wavefunction changes from the 2+ to the 3+ ground state since the Kondo effect becomes crucial for the Tm$^{2+}$ ions and is not strong enough for Tm$^{3+}$.

\begin{figure}
\begin{center}
\includegraphics[width=7cm]{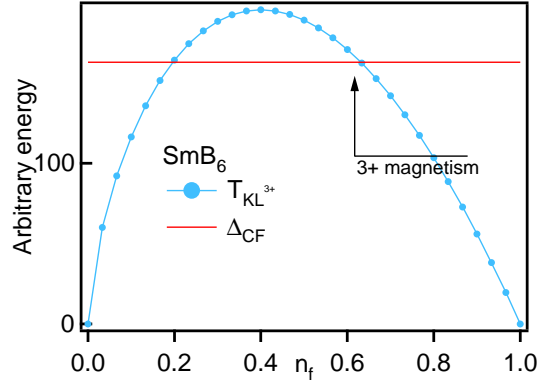}
\caption{Comparison of the Kondo lattice temperature and the crystal field spliting in arbitrary units, for the 3+ configuration. In the case of \smb\, as only the trivalent state is magnetic, long range magnetism will be allowed if T$_{KL}^{3+}$ become smaller than \cf.}
\label{simul_smb6}
\end{center}
\end{figure}    
For Sm Kondo lattice, the previous formula of \tkt\ is plotted in figure~\ref{simul_smb6}. Now, $\frac{\alpha D_0}{\Delta_{CF}}\sim 2$ is choosen. The interpretation is the same : a magnetic ground state is possible when its trivalent Kondo lattice temperature (in charge of the long range magnetism here, since Sm$^{2+}$ is non magnetic) is low enough compared to a crystal field energy. At low value of \nf\, the long range magnetism will disappear as the exchange energy will drop.

\paragraph{}The important point is that \tkt\ reaches a broad maxima near \nf $=0.4$. This is a critical difference with the Cerium case which correspond to the release of the 4f electron from the 4f shell : Ce$^{3+}\Longleftrightarrow$ Ce$^{4+}+5d$. If no extra electron is considered, one may find that \tkt\ goes, in Ce case, as \tkt $= (1-$\nf $)^{\frac{5}{3}} \alpha D_0 N_f$ (with $N_f=2J+1=6$). By contrast to the previous case, \tkt\ never reaches a maximum in the Ce case. Actually, this naive scheme gives the correct result that in Ce HFC, \tk\ decreases continuously with increasing \nf.

\paragraph{} In those considerations, there is the underlining assumption that the valence fluctuation can be slow enough to follow the motion of the spin dynamics of the trivalent configuration even for n$_f \sim 0.8$ as observed in SmS, \smb\ by NFS or in YbRh$_2$Si$_2$\cite{Sichelschmidt03}. This suggest that the 4f-5d correlation is a favorable factor to slow down the valence fluctuation. This consideration lead to propose that SmB$_6$, like SmS, can be regarded in the low pressure gold phase ($p<p_{\Delta}$) as an excitonic dielectric semiconductor with the electron promoted from f shells spread over the p orbitals of neighboring boron sites but with the same symmetry as the f electron in the central Sm site\cite{Kikoin1,Kikoin2}. An alternative idea is that the electron (5d) created by the mixing of the 4f state and the hole produced in the conduction band screen the 4f hole and form a bound state in a low carrier density medium\cite{Kasuya}. Up to now, there is no consideration on the pressure dependence of the 5d screening and thus on the disappearance of the reported many body effects.  In term of a Kondo approach, one may think that one way to describe the extra many body effect is to consider the possibility of the Kondo effect of the 5d electron itself. A many body treatment will be required, so far its \tk\ (5d) is lower than its crystal field splitting $\Delta_{CF}$ (5d). Of course, \tk\ (5d) will be far greater than \tk\ (4f) but also $\Delta_{CF}$ (5d) $>\Delta_{CF}$(4f). A change will occur under pressure since in  all reported cases (Sm$^{3+}$, Yb$^{3+}$, Tm$^{3+}$) their $T_K$(4f) decrease under $p$ while $T_K$(5d) increase with pressure. When $T_K$(5d)$>\Delta_{CF}$(5d) there will be no more reason to consider the extra many body effects of the 5d electron which could be considered then  as dissolved in the Fermi sea.

\paragraph{} In the case of TmSe, entering in the trivalent state, there are two reasons  that the physics will be dominated by the formation of a magnetic moment on an initial singlet ground state : the Kondo effect and a probable singlet crystal field level. As pointed out, the two mechanisms leads to rather similar increase of the sublattice magnetization under pressure on increasing the intersite exchange coupling. Thus the difference in the crystal field ground state limits the comparison of TmSe with SmB$_6$ and SmS. However, let's emphasize the similarity : up to \nf $\sim 0.8$, the physics appear renormalized to the divalent configuration, not only governing the magnetic properties, but also the electronic properties (formation of many body insulating state) ; above \nf $\sim 0.8$ the physics is now governed by the trivalent configuration (metallic conduction and nature of the magnetic order parameter). 

\paragraph{} Microscopic evidence to the 2+ configuration in SmS, \smb\ but even TmSe was given by inelastic neutron squattering experiments and measurement of the magnetic form factor\cite{MignotphysicaB,Alekseev2002}. The demonstration of a dressing towards the 3+ configuration for SmS and \smb\ was done by NFS as both the quadrupolar and dipolar magnetic hyperfine structure are caracteristic of a 3+ state even for \nf\ $\sim 0.8$. Macroscopically, suggestions of the 2+ renormalization of TmSe at $p=0$ comes from the specific heat, and of the 3+ renormalization of TmSe above 3~GPa from its continuity with the quasitrivalent compound TmS.

\paragraph{} Concerning SmS and \smb\, we have to be careful on the coincidence in the appearance of long range magnetism and closing of the hybridization gap. For SmS, the coincidence has been found. An extrapolation made from inelastic measurement on Sm$_{0.83}$Y$_{0.17}$S suggests strongly that Sm-Sm exchange interactions play a major role even in the low pressure gold phase\cite{Alekseev2002}. No similar influence is observed for \smb\ certainly due to the isolation of the Sm ion with the B cage. The gap is closed as function of pressure before long range magnetic ordering appears. Typically, the gap is closed between 4 and 6~GPa\cite{Gabani, Cooley, Moshchakov,Beille}, but long range magnetism do not appear before 8~GPa and a homogeneous magnetic phase picture without phase separation may occur only above 10~GPa. 

\paragraph{} Finally, the difference between \smb\ and SmS is not so surprising, as their band structures are completely different due to symmetries which are different. Local spin density approximation (LSDA+U approach) were published for Sm monochalcogenides\cite{Antonovsms} and \smb \cite{Antonovsmb}. For SmS, NaCl type  crystal structure with the space group Fm3m, the occupation number \nf\ is found equal to 0.55 (valence $v=2.55$) in the  low pressure gold phase, a non zero magnetic moment is always obtained.
For \smb\ , CaB$_6$ type crystal structure with the space group Pn3m, the calculations produce always an integer valence ground state either divalent or trivalent. A small hybridization energy gap is recovered in \smb\ for samarium in the divalent state. It was emphasized that the magnetism of golden SmS as well as the formation of the IV state in \smb\ requires to go beyond this mean field approximation.

\section{Conclusion}
\paragraph{} Ac calorimetry with in situ $p$ variation at low temperature is a powerful technique to define without ambiguity the magnetic phase diagram under pressure. We hope that our experimental report may help to experimental progresses.
\paragraph{} The common point in the three investigated systems TmSe, \smb , and SmS\cite{Haga,BarlaSmS} is the link between the electric conduction and the renormalization to divalent or trivalent configurations at low temperature. Looking more deeply on SmS and \smb , a main difference appear between the clear onset of antiferromagnetism at $p_{\Delta}$ in SmS and the large pressure window in \smb\ ($6<p<10$~GPa) where an inhomogeneous behaviour is observed. A homogeneous magnetic phase occurs in \smb\ only above 10~GPa. It is amazing to observe that if $p_{\Delta}=2$~GPa is remarkably reproducible in SmS\cite{Haga}, a large dispersion appears for \smb\ (around 3~GPa). The next step is to understand the role of the disorder in \smb\ and the impact on the collapse of the gap and the appearance of long range magnetism.
\paragraph{} Finally by comparison to results on Ce intermetallic heavy fermion compounds, in these Sm and Tm systems, a long range magnetism characteristic of the trivalent configuration occurs far  below the pressure where the trivalent state will be reached. This phenomena is quite similar to that observed in YbRh$_2$Si$_2$. Physically, the interesting fact is that both slow spin and valence fluctuations must interfer.

\section*{Aknowledgment}
\paragraph{} We would like to thank Christophe Marcenat for his precious clarification in the explanation of the phase behaviour of ac microcalorimetry measurements.


\begin{thebibliography}{99}
\bibitem{revuedeJacques} J. Flouquet, Cond-mat/0501602 (2005)
\bibitem{Wachter} P. Wachter, Handbook of physics and chemistry of rare earths, Edited by K.A. Gschneider {\it et al.} North holland Amsterdam (1994).
\bibitem{Malterre} D. Malterre Adv. Phys. {\bf 45} 299 (1996)
\bibitem{Launois} H. Launois, M. Rawiso, E. Holland-Moritz, R. Pott, and D. Wohlleben, Phys. Rev. Lett.  {\bf 44} 1271 (1980).  


\bibitem{Haen} P. Haen, F. Lapierre, J. M. Mignot, J. P. Kappler, G. Krill and M. F. Ravet, J. Magn. Magn. Mater.  {\bf 47-48}, 490 (1985).
\bibitem{Wertheim} G. K. Wertheim, W. Eib, E. Kaldis and M. Campagna, Phys. Rev. B {\bf 22}, 6240 (1980).
\bibitem{Brewer} W. D. Brewer, G. Kalkowski, G. Kaindl and F. Holtzberg, Phys. Rev. B {\bf 32}, 3676 (1985).
\bibitem{Beaurepaire} E. Beaurepaire, J. P. Kappler and G. Krill, Phys. Rev. B {\bf 41}, 6768 (1990).
\bibitem{Roeler} J. Röhler et al., in Valence Instabilities, edited by P.
Wachter and H. Boppart (North-Holland, Amsterdam,
1982), p. 215.
\bibitem{Dallera} C. Dallera, E. Annese, J-P Rueff, M. Grioni, G. Vanko, L. Braicovich, A. Barla, J-P. Sanchez, R. Gusmeroli, A. Palenzona, L. Degiorgi and G. Lapertot, J. Phys.: Condens. Matter {\bf 17} S849 (2005).

\bibitem{Ogita} N. Ogita, S. Nagai, M. Udagawa, F. Iga, M. Sera, T. Oguchi, J. Akimitsu and S. Kunii, Physica B {\bf 359-361} 941 (2005).


\bibitem{BarlaSmS} A. Barla, J. P. Sanchez, Y. Haga, G. Lapertot, B. P. Doyle, O. Leupold, R. Ruffer, M. M. Abd-Elmeguid, R. Lengsdorf and J. Flouquet, Phys. Rev. Lett.  {\bf 92} 066401 (2004).  
\bibitem{Haga} Y. Haga, J. Derr, A. Barla, B. Salce, G. Lapertot, I. Sheikin, K. Matsubayashi, N. K. Sato and J. Flouquet, Phys. Rev. B {\bf 70}, 220406 (2004)
\bibitem{BarlaSmB6} A. Barla, J. Derr, J. P. Sanchez, B. Salce, G. Lapertot, B. P. Doyle, R. Ruffer, R. Lengsdorf, M. M. Abd-Elmeguid and J. Flouquet, Phys. Rev. Lett. {\bf 94}, 166401 (2005).
\bibitem{Ribault} M. Ribault, J. Flouquet, P. Haen, F. Lapierre, J. M. Mignot and F. Holtzberg, Phys. Rev. Lett.  {\bf 45} 1295 (1980).  
\bibitem{Ohashi1} M. Ohashi, N. Takeshita, H. Mitamura, T. Matsumura, T. Suzuki, T. Goto, H. Ishimoto and N. Môri, Physica B {\bf 259-261}, 326-328 (1999).
\bibitem{Ohashi2} M. Ohashi, N. Takeshita, H. Mitamura, T. Matsumura, T. Suzuki, T. Mori, T. Goto, H. Ishimoto and N. Môri, J. Magn. Magn. Mater.  {\bf 226-230}, 158-160 (2001).

\bibitem{TmSeTmSauCNRS} F. Holtzberg, J. Flouquet, P. Haen, F. Lapierre, Y. Lassailly and C. Vettier, J. of Appl. Phys. {\bf 57} (1985) 3152-3153.
\bibitem{bernard} B. Salce, J. Thomasson, A. Demuer, J. J. Blanchard, J. M. Martinod, L. Devoile and A. Guillaume, Rev. sci. Instrum. {\bf 71}, 2461 (2000).
\bibitem{albin} A. Demuer, C. Marcenat, J. Thomasson, R. Calemczuk, B. Salce, P. Lejay, D. Braithwaite and J. Flouquet, J. of low Temp Phys, {\bf 120} 245-57 (2000)
 

\bibitem{accalorimetry} Paul F. Sullivan and G. Seidel, Phys. Rev. {\bf 173}, 679-685 (1968).
\bibitem{chrisetmam} C. Marcenat and M. A. Méasson, private communication. 
\bibitem{Berton} A. Berton, J. Chaussy, B. Cornut, J. Flouquet, J. Odin, J. Peyrard and F. Holtzberg, Phys. Rev. B, {\bf 23} 3504 (1981).
\bibitem{TmSPzero} A. Berton, J. Chaussy, J. Flouquet, J. Odin, J. Peyrard and F. Holtzberg, Phys. Rev. B, {\bf 31} 4313 (1985).

\bibitem{CeB6} T. Fujita, M. Suzuki, T. Komatsubara, S. Kunii, T. Kasuya and T. Ohtsuka, Solid-State-Communications. Aug. 1980; 35(7): 569-72
\bibitem{Shiina98} R. Shiina, O. Sakai, H. Shiba and P. Thalmeier, J. Phys. Soc. Jpn. {\bf 67} 941 (1998). 
\bibitem{TheseMAM} M. A. Méasson, Thèse de doctorat, Université Joseph Fourrier, Grenoble (2005). 

\bibitem{jean} J. Thomasson {\it et al.} Private communication.


\bibitem{Mignot} J. M. Mignot, I. N. Goncharenko,P. Link,T. Maysumura and T. Suzuki, Hyperfine interactions {\bf 128} (2000) 207-224.


\bibitem{OashiTmS} M. Ohashi, N. Takeshita, H. Mitamura, T. Matsumura, Y. Uwatoko, T. Suzuki, T. Goto, H. Ishimoto and N. Môri, Physica B {\bf 281-282}, 264-266 (2000).

\bibitem{TmSdefect} G. Chouteau, F. Holtzberg, O. Pena, T. Penney and R. Tournier, J. Physique, colloq 40 (1979) C5-361. 
\bibitem{compress1} Y. Lassailly, C. Vettier ,F. Holtzberg, J. Flouquet, C. M. E. Zeyen and F. Lapierre, Phys. Rev. B 28, 2880-2882 (1983)
\bibitem{compress2} B. Batlogg, H. R. Ott, E. Kaldis, W. Thöni, and P. Wachter, Phys. Rev. B {\bf 19}, 247-259 (1979).

\bibitem{Gabani} S. Gabani, E. Bauer, S. Berger, K. Flashbart, Y. Paderno, C. Paul, V. Pavl\'ik and N. Shitsevalova, Phys. Rev. B, {\bf 67} 172406 (2003).
\bibitem{Moshchakov} V.V. Moshchalkov, I. V. Berman, N.B. Brandt, S. N. Pashkevich, E.V. Bogdanov, E. S. Konovalova and M. V. Semenov, J. Magn. Magn. Mater.  {\bf 47-48}, 289-291 (1985).
\bibitem{Cooley} J. C. Cooley, M. C. Aronson, Z. Fisk and P. C. Canfield, Phys. Rev. Lett.  {\bf 74} 9 (1995).  
\bibitem{Beille} J. Beille, M. B. Maple, J. Wittig, Z. Fisk and L. E. Delong, Phys. Rev. B, {\bf 28} 12 (1983).


\bibitem{argon} K. V. Tretiakov and S. Scandolo, Journal of Chemical Physics. (2004) 121(22) 11177-82
\bibitem{argonbis} H. Shimizu, H. Tashiro, T. Kume, S. Sasaki, Phys. Rev. Lett. {\bf 86} 20, 4568-71 (2001)

\bibitem{Flouquet} J. Flouquet, A. Barla, R. Boursier, J. Derr and G. Knebel, J. Phys. Soc. Jpn. {\bf 74} 178-185 (2005). 

\bibitem{Hewson} A. C. Hewson, The kondo problem to heavy fermions (Cambridge University Press, Cambridge, 1993).
\bibitem{Newns} D.M. Newns and N. Read, Adv in Physcics, {\bf 36(6)} 799-849 (1987)
\bibitem{Read} N. Read, K. Dharamvir, J. W. Rasul and D. M. Newns, J. Phys. C {\bf 19} (1986) 1597.
\bibitem{Nunes} A. C. Nunes, J. W. Rasul and G. A. Gehring, J. Phys. C {\bf 19} (1986) 1017.
\bibitem{Yafet} Y. Yafet, C. M. Varma  and B. Jones, Phys. Rev. B, {\bf 32} 360 (1985).
\bibitem{Saso} T. Saso, J. Magn. Magn. Mater.  {\bf 76-77}, 176-178 (1988).

\bibitem{Sakai} O. Sakai, Y. Shimizu and T. Kasuya, Prog. Theor. Physics, {\bf 108} 73 (1992).

\bibitem{reforbital} J. T. Waber and D. T. Cromer, J. Chem. Phys. {\bf 42} (1965) 4116.
\bibitem{Harima} H. Harima, private communication. 
\bibitem{Plessel2003} J. Plessel, M. M. Abd-Elmeguid, J. P. Sanchez, G. Knebel, C. Geibel, O. Trovarelli and F. Steglich,  Phys. Rev. B, {\bf 67} 180403 (2003).
\bibitem{Knebel2005} G. Knebel, V. Glaskov, A. Pourret, P. G. Nicklowitz, G. Lapertot, B. Salce and J. Flouquet, Physica B {\bf 359-361} (2005) 20-22.

\bibitem{Sichelschmidt03} J. Sichelschmidt, V. A. Ivanshin, J. Ferstl, C. Geibel, and F. Steglich, Phys. Rev. Lett.  {\bf 91} 156401 (2003).  
\bibitem{Kikoin1} K. A. Kikoin and A. S. Mishchenko, J. Phys. cond. matt. {\bf 7} 307 (1995).
\bibitem{Kikoin2} S. Curnoe and K. A. Kikoin, Phys. Rev. B, {\bf 61} 15714 (2000).
\bibitem{Kasuya} T. Kasuya, Europhysics letter {\bf 26} 277 (1994).




\bibitem{MignotphysicaB} J. M. Mignot and P. A. Alekseev, Physica B {\bf 215}, 99 (1995).
 

\bibitem{Alekseev2002} P. A. Alekseev, J.-M. Mignot, A. Ochiai, E. V. Nefeodova, I. P. Sadikov, E. S. Clementyev, V. N. Lazukov, M. Braden, and K. S. Nemkovski, Phys. Rev. B, {\bf 65} 153201 (2002).
 
\bibitem{Antonovsms} V. N. Antonov, B. N. Harmon and A. N. Yaresko Phys. Rev. B, {\bf 66} 165208 (2002).
\bibitem{Antonovsmb} V. N. Antonov, B. N. Harmon and A. N. Yaresko Phys. Rev. B, {\bf 66} 165209 (2002).



\end{thebibliography}
\end{document}